# Quadrature magnetoresistance scaling reflects linear field dependence rather than strange metallicity of high-temperature superconductors

Dingbang Zhou,[1,2,*] Yang Yang,[1,*] Liefeng Feng,[1] Mengfan Zhao,[3] Zhiyan Jia,[3] and Kuanghong Gao[1,†]

[1]Tianjin Key Laboratory of Low Dimensional Materials Physics and Preparing Technology, Department of Physics, Tianjin University, Tianjin 300354, China

[2]Analysis and Testing Center of Tianjin University, Tianjin 300072, China

[3]Institute of Quantum Materials and Devices, Tiangong University, Tianjin 300387, China

[*]These authors have equal contributions to this work.

[†]Author to whom any correspondence should be addressed.

**E-mail:** khgao@tju.edu.cn

## Abstract

The quadrature scaling of magnetoresistance (MR)—the collapse of resistivity data as a function of *B*/*T*—has been widely adopted as a hallmark of the strange-metal state in high-temperature superconductors. However, whether this scaling signals quantum criticality or reflects conventional transport behavior remains controversial. Here, by systematically investigating the magnetotransport properties of $NiTe_2$ nanosheets, we demonstrate that the quadrature scaling is not a unique signature of strange metallicity. We find that the scaling holds only when the crossover field $B_c$, marking the transition from quadratic to linear MR, is sufficiently small relative to the applied field range. Through controlled simulations, we show that the scaling emerges whenever linear MR dominates, irrespective of its origin, and fails when the linear regime is inaccessible. This conclusion is supported by observations in $Al_2O_3/SrTiO_3$ heterostructures, where quadrature scaling appears despite the strange-metal state has never been reported before. Our results establish that the quadrature scaling merely reflects the presence of linear MR, urging caution in using this scaling as a diagnostic tool for exploring the strange-metal state.

## 1. Introduction

In conventional metals, resistivity is expected to exhibit a quadratic temperature dependence owing to electron-electron scattering accompanied with a positive quadratic magnetoresistance (MR) for which Kohler's rule is valid [1]. However, the normal-state resistivity in some high-temperature superconductors shows a linear temperature dependence together with a nonsaturating linear MR (resistivity increases linearly with external magnetic field), suggesting a non-Fermi-liquid behavior. This indicates that conventional quasiparticle concept breaks down and there emerges a strange-metal state [2]. For the strange-metal state, a new scaling behavior for MR is firstly found in iron-based superconductor near its quantum critical point [3]:

$$[\rho_{xx}(B,T)-\rho_{xx}(0,0)]/T=\alpha\sqrt{1+(\gamma B/\alpha T)^2}\,, \quad (1)$$

where $\alpha$ and $\gamma$ are two parameters and $\rho_{xx}(0,0)$ is the extrapolated zero-field residual resistivity. Here, we refer to it as the quadrature scaling. The quadrature scaling is proposed as a potential typical character of the strange-metal state although whose origin is still under intense debate [4]. From then on, it has been widely used to study the MR for exploring possible strange-metal state in various systems including cuprate superconductors [5-7], pnictide superconductors [8], FeSe [9, 10], and CrAs [11]. In pnictide superconductors, the quadrature scaling is found to compete with the Kohler scaling, signifying a non-conventional nature of the strange-metal state [8]. In $LaNiGa_2$ as a candidate topological superconductor, however, the quadrature scaling is recently reported to coexist with the Kohler scaling in the higher temperatures [12]. This implies that the connection between these scaling behaviors remains controversial.

Theoretically, the quadrature scaling is usually viewed as a typical character of the strange-metal state. Patel et al. consider a macroscopically disordered sample with domains of diffusive marginal-Fermi liquid and find that the conductivity satisfies the quadrature scaling [13]. Hinlopen et al. think that the quadrature scaling arises from the impeded orbital motion and can be explained by a modified Boltzmann theory [14]. Kim et al. construct a microscopic model by coupling electronic excitations to quantum critical bosons via a spatially disordered Yukawa interaction and reproduce the quadrature scaling [15]. In contrast to these reports, Singleton maintains that the quadrature scaling is not a definitive signature of the strange-metal state, but rather arises solely from the disorder [16]. This indicates that whether the quadrature scaling is a specific character of the strange-metal state is still an open question.

$NiTe_2$ as one of typical Dirac semimetals is of interest because it reveals some interesting phenomena such as spin-polarized surface state and Lifshitz transition [17,18]. In this work, we investigate the magnetotransport properties of $NiTe_2$ nanosheets, and observe two MR components at low temperatures: quadratic field dependence at low fields and linear field dependence at high fields. By investigating nanosheets with varying $B_c$, which is a crossover field from quadratic to linear MR, we surprisingly find that the scaling holds only when $B_c$ is sufficiently small relative to the applied field range. This indicates that the quadrature scaling is not a hallmark of strange metallicity but merely reflects the dominance of linear MR.

## 2. Experimental details

$NiTe_2$ single crystal was grown by chemical vapor transport method. Ni (99.999%) and Te (99.999%) powders with a 1:5 ratio were mixed and sealed in a quartz tube under high vacuum. The temperature of the quartz tube was raised up to 973 K in a furnace. Then it was maintained for 4 days. After that, the quartz tube was cooled to room temperature slowly. The grown $NiTe_2$ single crystals were characterized by transmission electron microscopy, X-ray diffraction, and energy dispersive X-ray spectroscopy measurements, and found to be of high quality [19]. By using mechanical exfoliation, $NiTe_2$ nanosheets were obtained. The exfoliated nanosheets were transferred onto silicon substrates with a ~300 nm $SiO_2$ cover layer. The thickness of the exfoliated nanosheets was determined by atomic force microscopy. A standard photolithography was used to fabricate six Ohmic contacts on single nanosheet [see the inset of Fig. 1(d)]. Six nanosheets are referred to as samples #1-#6 with the thickness increasing from 52.3 to 202.8 nm in sequence (Table 1). The physical property measurement system (Quantum Design) was used to perform the magnetotransport measurement with a magnetic field perpendicular to the plane of the sample, and longitudinal (Hall) data was symmetrized (antisymmetrized) about zero magnetic field.

**3. Results and discussion**

Figure 1(a) shows the normalized resistivity $\rho_{xx} / \rho_{xx}(300\,\mathrm{K})$ as a function of temperature for all the studied samples with different thicknesses. One can see that $\rho_{xx} / \rho_{xx}(300\,\mathrm{K})$ decreases nearly linearly on decreasing temperature from 300 K down ~20 K for any given sample, indicating a metallic character. Below ~20 K, $\rho_{xx} / \rho_{xx}(300\,\mathrm{K})$ is saturated to a residual resistivity and no superconducting transition occurs down to 2 K. This is in agreement with observation that the superconducting transition temperature in $NiTe_2$ is lower than 0.3 K [20]. Figure 1(b) shows the resistivity $\rho_{xx}$ as a function of field at some selective temperatures for sample #4. At 2 K, $\rho_{xx}$ exhibits a positive quadratic field dependence in the low field range of 0 - ~2 T, while it is translated into a linear field dependence above ~2 T and no saturating trend emerges up to 9 T. This means that there are two MR components: quadratic MR at low fields and linear MR at high fields, consistent with the reported results in other superconductors with the appearance of the strange-metal state [6-10, 21]. As temperature increases, one can see that the quadratic MR is gradually extended to the higher field and the linear MR completely disappears at 100 K (corresponds to a case that $B_\mathrm{c} > 9\,\mathrm{T}$ ). For convenience, we further define the MR as $\mathrm{MR} = [\rho_{xx}(B,T) - \rho_{xx}(0,T)] / \rho_{xx}(0,T) \times 100\%$. In other samples, as seen in Fig. 1(c), the linear MR is also evident in the high fields at 2 K. Particularly, sample #1 seems to show the quadratic MR over through the whole field range (see the inset), indicating a higher $B_\mathrm{c}$ (discussed later). Figure 1(d) shows the quadrature scaling of MR for sample #2. Clearly, one can see that all $[\rho_{xx}(B,T) - \rho_{xx}(0,0)] / T$ data in the temperature range of 2-20 K can be scaled into a single line. Above 20 K, however, a large deviation appears. This deviation is understandable because $\rho_{xx}(0,T)$ starts to increase with temperature above ~20 K [see Fig. 1(a)], which makes $\rho_{xx}(B,T)$ larger than residual resistivity and consequently causes $[\rho_{xx}(B,T) - \rho_{xx}(0,0)] / T > 0$. In order to reconcile this deviation at high temperatures, it is necessary to introduce an additional temperature-dependence term into $\rho_{xx}(0,0)$, as proposed in cuprate superconductors [6]. We tentatively substitute $\rho_{xx}(0,T)$ for $\rho_{xx}(0,0)$ to check the quadrature scaling.

As shown in Fig. 2, we can find that the data for $[\rho_{xx}(B,T)-\rho_{xx}(0,T)]/T$ at various temperatures collapse onto a single scaling function for samples #2, #3, and #6, indicating that the quadrature scaling is valid. In contrast, the MR in other samples (i.e., #1, #4, and #5) does not exhibit the well-developed quadrature scaling especially in sample #1. There seems to exist the quadrature scaling above ~8 K for samples #4 and #5, which will be discussed later. Considering quasi-linear temperature-dependent resistivity at high temperatures [Fig. 1(a)] and the linear MR at high fields [Fig. 1(c)] combined with the well-developed quadrature scaling (Fig. 2) as three characters of samples #2, #3, and #6, the strange-metal state appears to be present in these samples.

However, the strange-metal state is likely absent in all our samples due to five following reasons. First, we fitted $\rho_{xx}$ vs $T$ plots by the power law $\rho_{xx}=A+BT^n$ ($A$, $B$, and $n$ are fitting parameters) in the temperature range from 2 K up to ~30 K. The reason for adopting this fitting window can be found in the section 1 of supplementary material. As seen in Fig. 1(a), the experimental data can be well described for all studied samples (dashed lines are fits). The fit-obtained $n$ varies between 2.26±0.02 and 2.62±0.02 (Table 2). Namely, $n$ is larger than 2, which might be related to disorder, strain, multiband effect, or umklapp scattering that need further study [22-25]. Importantly, these $n$ values deviate from the relation $n<2$ (specifically, $n\approx 1$) that is believed to be related to the strange-metal state [26,27]. Second, we fit the $\rho_{xx}$ vs $T$ plots at high temperatures by using the Bloch-Grüneisen formula [28]:

$$\rho_{xx}(0,T)=\rho_{xx}(0,0)+MT\left(T/\theta_{\mathrm{D}}\right)^4\int_0^{\theta_{\mathrm{D}}/T}\frac{x^5\mathrm{d}x}{(e^x-1)(1-e^{-x})}, \tag{2}$$

where $M$ is a material constant, and $\theta_{\mathrm{D}}$ is the Debye temperature. In a wide temperature range of ~30 – 300 K, $\rho_{xx}$ vs $T$ plots can be well reproduced by Eq. (2) (see section 2 in supplementary material). This means that the quasi-linear temperature dependence of resistivity can be explained by the electron-phonon scattering. Interestingly, the extracted $\theta_{\mathrm{D}}$, as seen in Table 2, roughly increase on increasing thickness, which can be attributed to the surface-caused phonon softening and similar results were reported in gold, copper, and platinum films [29-31]. Third, we find that the MR satisfies the extended Kohler scaling [32]. As seen in Fig. 3(a) for sample #6, MR vs $B/\rho_{xx}(0,T)$ plots at various temperatures do not collapse onto a single curve. This hints that the Kohler scaling is broken down. However, when an additional scaling factor $n_{\mathrm{T}}$ was introduced with setting $n_{\mathrm{T}}=1$ at 2 K, we found that all curves at various temperatures can be scaled onto a single curve by adjusting $n_{\mathrm{T}}$ [see Fig. 3(b)]. The extracted $n_{\mathrm{T}}$, as seen in Fig. 3(c), increases as temperature increases, which is the same trend as the temperature dependence of total carrier concentration $N_{\mathrm{T}}$ obtained by Hall analysis (section 3 in supplementary material). This suggests that the MR data for our samples follows the extended Kohler's rule, signifying a conventional transport behavior again. The similar results were also obtained in other samples and Fig. 3(d) shows another example of sample #1.

Fourth, a prevailing speculation about the linear temperature-dependent resistivity of the strange metal state involves the Planckian dissipation [33] with the scattering rate given by $\hbar/\tau=\alpha' k_{\mathrm{B}}T$ ($\hbar$ is the reduced Planck constant, $\alpha'$ is a constant, and $k_{\mathrm{B}}$ is the Boltzmann constant). According to the reported results, $\alpha'$ should be close to unity for the strange-metal state (i.e., $\alpha'$ is reported to vary between 0.5 and 1.5) [21, 33-35]. To test this point in our samples, we estimate $\alpha'$ by using the following formula [34]:

$$\alpha'=\frac{Ne^2\hbar k_{\mathrm{F}}^2}{2\pi c k_{\mathrm{B}}}\sum_i(1/m_i^*)\frac{d\rho_{xx}(0,T)}{dT}, \tag{3}$$

where $N$ is the number of formula units per unit cell, $k_{\mathrm{F}}$ is the Fermi wave vector, $c$ is the $c$-axis lattice parameter, $m^*$ is the effective mass, $d\rho_{xx}(0,T)/dT$ is the slope of the linear temperature-dependent resistivity. For $NiTe_2$, one can find that $N=2$ [34], $c=0.5260\ \mathrm{nm}$ [19]. Because of multiband nature, two effective masses ($m_1^*=0.24\,m_0$ and $m_2^*=0.31\,m_0$) [36] and two Fermi wave vectors ($k_{\mathrm{F1}}=0.15$ Å$^{-1}$ and $k_{\mathrm{F2}}=0.23$ Å$^{-1}$) [37] have been reported. Substituting $k_{\mathrm{F1}}$ into Eq. (3), we calculated $\alpha'$ that varies between 1.8 and 5.4 for all studied samples. If using $k_{\mathrm{F2}}$, the calculated $\alpha'$ value is in the range of 4.1-12.7. If substituting the total Fermi wave vector $k_{\mathrm{FT}} = (k_{\mathrm{F1}}{}^2 + k_{\mathrm{F2}}{}^2)^{0.5}$ into Eq. (3), as listed in Table 2, the calculated $\alpha'$ value is in the range of 5.9-18.2. In all cases, $\alpha'$ is substantially larger than 1.5, indicating the absence of the strange-metal state. Last, Shi et al. [7] recently studied the correlation between linear temperature-dependent resistivity and linear MR for various materials showing the strange-metal state. They found a fixed ratio ~ 0.40 of the $d\rho_{xx}(0,T)/dT$ to the slope of the linear MR [i.e., $d\rho_{xx}(B,T)/dB$] at a given temperature. For comparison, we calculated two parameters $\eta$ and $\beta$ by using the relations $\eta = [d\rho_{xx}(0,T)/dT]/k_{\mathrm{B}}$ and $\beta = [d\rho_{xx}(B,T)/dB]/\mu_{\mathrm{B}}$ at 2 K (here, $\mu_{\mathrm{B}}$ is the Bohr magneton), respectively. Then, the ratio $\beta/\eta$ can be obtained. As seen in Table 2, the obtained $\beta/\eta$ shows a large deviation from the reported value of 0.40 for all our samples. We should emphasize that the Planckian benchmark ($\alpha' \approx 1$) and the ratio of $\beta/\eta \approx 0.40$ are two empirical correlations; the present data fall outside the ranges reported for strange-metal state. It can be concluded that the present evidence does not support a strange-metal state although the well-developed quadrature scaling is observed in some samples like #2, #3, and #6.

Now let us turn to an important question: why does the quadrature scaling hold in some samples but fail in others (see Fig. 2)? The aforementioned observation that there are two MR components (i.e., quadratic MR at low fields and linear MR at high fields) at low temperatures is reminiscent of an empirical formula [38]

$$\mathrm{MR} = \sqrt{a^2 + (kB)^2} - a\,, \qquad (4)$$

where $a$ and $k$ are two parameters independent on field, and $B_{\mathrm{c}} = a/k$. Although the equation is empirical in nature, Maksimovic et al. obtained a similar expression that has been used to explain the MR in pnictide superconductors [8]. If the first term on the right-hand side of Eq. (4) is much larger than the second term under a large field, the second term can be neglected and Eq. (4) is simplified to

$$[\rho_{xx}(B,T) - \rho_{xx}(0,T)] = \rho_{xx}(0,T)\sqrt{a^2 + (kB)^2}\,. \qquad (5)$$

One can find that Eq. (5) has the completely same form as the Eq. (1) of the quadrature scaling. Furthermore, if $kB \gg a$ (corresponding to $B \gg B_{\mathrm{c}}$ due to $B_{\mathrm{c}} = a/k$), Eq. (5) is approximate to the expression $[\rho_{xx}(B,T) - \rho_{xx}(0,T)] \approx [\rho_{xx}(0,T)k]B$. We can find that the linear MR appears. This implies that the quadrature scaling possibly only captures linear MR component at high fields $B \gg B_c$. In order to check this conjecture, we fitted the MR curves by using Eq. (4). As seen in Fig. 1(c), the MR curves can be well reproduced for different samples including #1 in the inset (dashed lines are fitting curves).

The fit-obtained $B_c$ at 2 K is given in Table 2. One can see that sample #1 has a maximum $B_c$ of 7.67±0.07 T, which explains why the quadratic MR seems to extend over through the whole field range [see the inset of Fig. 1(c)]. It is worthy noting that $B_c$ does not correlate with the thickness of the nanosheets. This can be explained by the observation that $B_c$ is controlled dominantly by the classical disorder [39] while the multiband effect becomes important to determine $B_c$ in high-mobility sample like #5 (section 3 in supplementary material). Regarding the specific case of sample #3, a small $N_T$ (see Table 1) must be main reason for a relatively large residual resistivity while a small $B_c$ results from the high $\mu_E$ (the larger $B_c$ in samples #4-#6 with the higher $\mu_E$ can be attributed to multiband effect). Note that the residual resistance ratio RRR [$= \rho_{xx}(300\,\mathrm{K}) / \rho_{xx}(2\,\mathrm{K})$] is not related to thickness (Table 1). This implies that the disorder does not correlate with the thickness presumably due to uncontrollable surface contamination and formation of defects revealing a sample-to-sample variation [19]. As a consequence, there is no a clear thickness dependence of $B_c$. Importantly, we find that the quadrature scaling holds in samples #2, #3, and #6 with a small $B_c$ value (i.e., $B_c < 1.2\,\mathrm{T}$) at 2 K while the scaling fails in others with the larger $B_c$ value (i.e., samples #1, #4, and #5 with $B_c > 2.3\,\mathrm{T}$). The larger $B_c$ is indicative of a relatively narrower field range in which the linear MR persists. The narrower field range consequently destroys the quadrature scaling in samples #1, #4, and #5. This supports our conjecture: the quadrature scaling only captures linear MR component at high fields.

To confirm this point, we further fit the MR data at various temperatures by using Eq. (4). Figure 4(a) shows the temperature dependence of the fit-obtained $a$ and $k$ for sample #5. The former is roughly temperature independent while the fit-obtained $k$ is also temperature independent in the low temperature range of 2-20 K but shows a monotonous decrease on increasing temperature from 20 K up to 100 K. This causes a saturated $B_c$ at low temperatures and makes it increase with temperature above 20 K, as seen in Fig. 4 (d) (closed triangles). Based on the fit-obtained $a$ values, we can get new data by using the relation $a' = 10a$ while the fit-obtained $k$ is unchanged. Then, a new crossover field $B_c'$ can be calculated, and apparently $B_c' = a'/k = 10B_c$. As seen in Fig. 4(d) (open squares), $B_c'$ exhibits the same temperature dependence as $B_c$ but shifts up vertically. Note that $B_c' > 20\ \mathrm{T}$ over the whole studied temperature range. By using the calculated $a'$ and $k$, we simulated the MR data via Eq. (4) in the field range of 0-9 T. As shown in Fig. 4(b), there appears a series of quadratic MR curves for selective temperatures, which is expected because $B_c'$ is higher than 9 T that is the maximum field we can realize for any selective temperature. Figure 4 (c) shows the quadrature scaling of the simulated MR at various temperatures. One can see that all the curves do not collapse onto a single line, indicating the invalidity of the quadrature scaling. In contrast, if we use a lower crossover field $B_c'' = 0.1B_c$ [open circles in Fig. 4(d)], the simulated MR curves at various temperatures exhibits a clear linear field dependence at high fields because $B_c'' < 1.5\ \mathrm{T}$ for any selective temperature [see Fig. 4(e)]. Obviously, as seen in Fig. 4(f), the quadrature scaling is valid. This, together with the invalidity of the quadrature scaling for $B_c' = 10B_c$, confirms that the quadrature scaling only captures linear MR component.

Following this idea, if the external field is so high ($B \gg B_c'$) that the linear MR appears in a wider high field range, the quadrature scaling should be valid in high $B/T$ range. In order to prove it, we further simulated MR curves at various temperatures with field reaching to 1000 T (i.e., far higher than maximum $B_c'$ of 140 T at 100 K) when $B_c' = 10B_c$. Expectedly, as seen in the inset of Fig. 4(c), the quadrature scaling is well developed. Meanwhile, considering that $B_c'' = 0.3\ \mathrm{T}$ at 2 K (i.e., only quadratic MR component remains in the low fields less than 0.3 T), one can expect a large deviation between the quadrature scaling plot at 2 K and those plots at high temperatures (e.g., 100 K) that is present in low $B/T$ range. Indeed, as seen in the inset of Fig. 4(f), no any collapsing trend emerges between the quadrature scaling plots at 2 K and 100 K. The phenomenon cannot be clearly observed in the main panel of Fig. 4(f) because of low $B/T$ range for high-temperature plots. It should be mentioned that the approximate quadrature scaling appears above ~8 K for both samples #4 and #5 in Fig. 2. This can be explained by the low-$B/T$-induced window effect (section 4 in supplementary material).

Now, we can arrive at a clear conclusion: the quadrature scaling only captures the linear MR character, and for prevailing transport system where both quadratic and linear MRs coexist it is valid exactly when there is a small $B_c$ value. Empirically, it should be comparable to or less than about one tenth of the maximum applied field. In other words, the quadrature scaling of MR either does not reflect new physics or does not signify the emergence of a new transport state, and it is only suggestive of presence of linear field-dependent component dominating MR without revealing its origin. Naturally, the quadrature scaling is likely valid when the linear MR occurs regardless of its origin.

In addition to the strange-metal state, there are many proposed mechanisms for the linear MR, including classical disorder [39,40], quantum model with the presence of Dirac fermions [41], guiding center diffusion [42], spin/charge density wave fluctuations [43,44], electron-hole compensation [45], and Berry curvature [46]. Recently, we observed a disorder-induced linear MR in a series of $Al_2O_3/SrTiO_3$ heterostructures [47] in which no the strange-metal state has been reported before. For one sample with a smaller $B_c$ (< 0.1 T) at low temperatures, as seen in Fig. 5, a well-developed quadrature scaling of MR is observed although the strange-metal state is not found in the $SrTiO_3$-based materials. This indicates that we need to reconsider the quadrature scaling as a reliable tool to study the strange-metal state.

**4. Conclusion**

We study the quadrature scaling of MR in $NiTe_2$ nanosheets with different $B_c$. The quadrature scaling is satisfied in some samples although in which no the strange-metal state is present. Inspecting two MR components (quadratic and linear field dependences), we find that the quadrature scaling is only valid for MR curves with a small $B_c$. Via data simulation, we confirm that the quadrature scaling of MR only reflects the presence of linear MR but not signifies the specific transport property of the strange-metal state. Therefore, a cautious approach should be adopted when applying this scaling relation to explore the strange-metal state.

**Acknowledgement**

This work was supported by National Natural Science Foundation of China (No. 12274319), National Key Research and Development Program (No. 2024YFA1208400), and Frontier Fundamental Research Program of Tianjin University (No. 2025XJ21-0005).

**Data availability statement**

All data that support the findings of this study are included within the article (and any supplementary files).

**Conflict of interest**

The authors have no conflicts to disclose.

**Table 1.** Basic parameters of the exfoliated nanosheets. RRR is the residual resistance ratio. $N_{\mathrm{T}}$ and $\mu_{\mathrm{E}}$ are respectively total carrier concentration and effective mobility.

| Sample | #1 | #2 | #3 | #4 | #5 | #6 |
|---|---|---|---|---|---|---|
| Thickness (nm) | 52.3 | 64.6 | 113.2 | 165.6 | 182.7 | 202.8 |
| RRR | 1.8 | 21.4 | 6.8 | 15.9 | 14.5 | 22.4 |
| $N_{\mathrm{T}}$ ($10^{22}$ cm$^{-3}$) | 25.0 | 6.0 | 2.6 | 3.6 | 5.2 | 2.0 |
| $\mu_{\mathrm{E}}$ (cm$^{2}$V$^{-1}$s$^{-1}$) | 29 | 103 | 224 | 1399 | 1200 | 2670 |

**Table 2.** $n$ is the exponent of power law. $\theta_{\mathrm{D}}$ is the Debye temperature. $\alpha'$ is a constant related to the Planckian dissipation. $\eta$ and $\beta$ is defined as $\eta = [d\rho_{xx}(0,T)/dT]/k_{\mathrm{B}}$ and $\beta = [d\rho_{xx}(B,T)/dB]/\mu_{\mathrm{B}}$, respectively. $B_{\mathrm{c}}$ is the crossover field.

| Sample | #1 | #2 | #3 | #4 | #5 | #6 |
|---|---|---|---|---|---|---|
| $n$ | 2.26±0.02 | 2.47±0.02 | 2.38±0.03 | 2.52±0.01 | 2.56±0.02 | 2.62±0.02 |
| $\theta_{\mathrm{D}}$ (K) | 160±8 | 194±1 | 203±9 | 236±1 | 225±2 | 192±4 |
| $\alpha'$ | 7.1±0.1 | 7.3±0.3 | 7.3±1.5 | 18.2±0.1 | 10.0±0.1 | 5.9±0.1 |
| $\beta/\eta$ | 0.22±0.01 | 1.69±0.05 | 1.33±0.09 | 1.36±0.03 | 1.44±0.04 | 1.49±0.03 |
| $B_{\mathrm{c}}$ (T) | 7.67±0.07 | 0.71±0.01 | 0.77±0.01 | 2.32±0.02 | 2.70±0.03 | 1.18±0.01 |

Figure 1

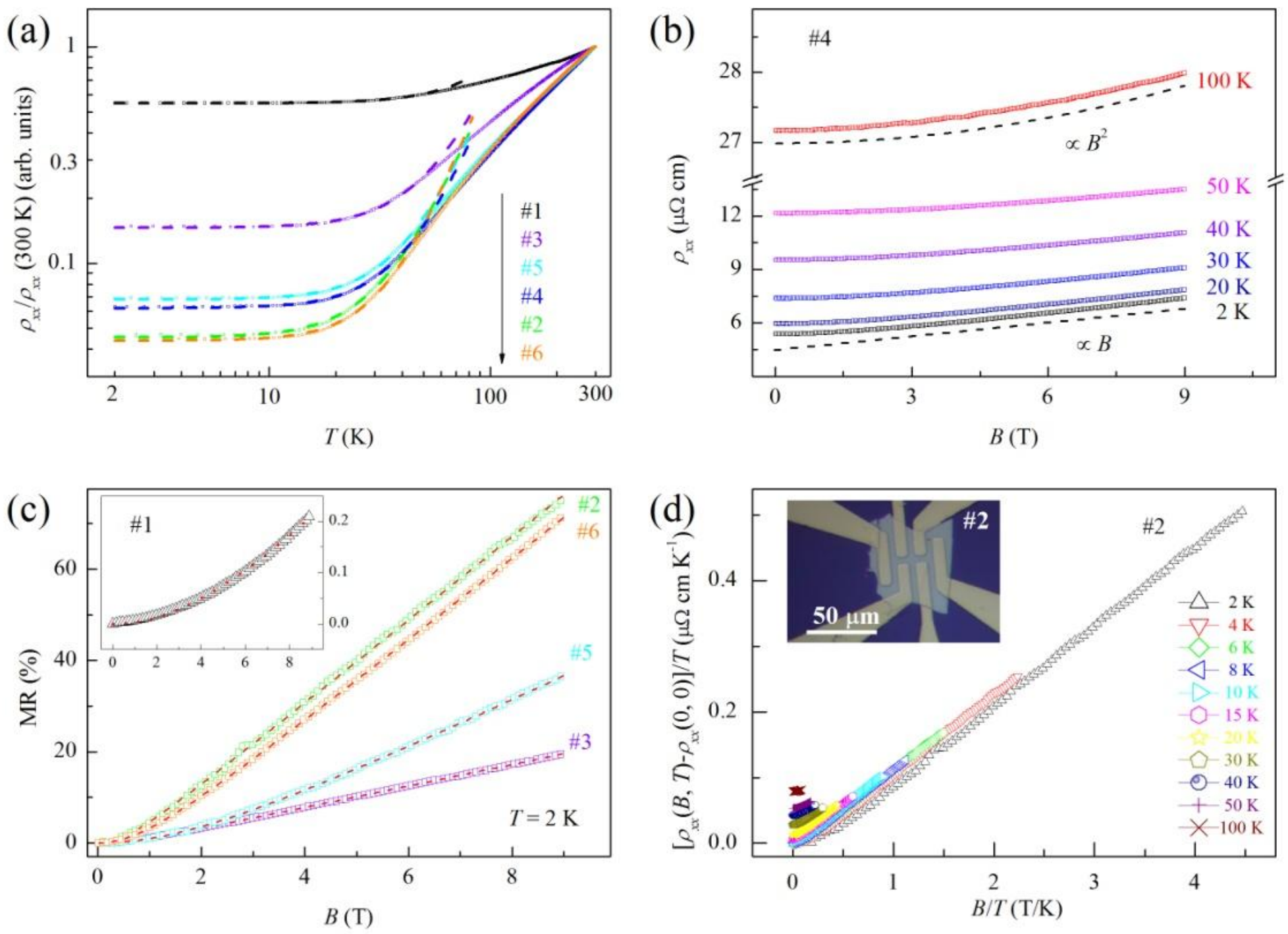

**Figure 1** (a) Normalized resistivity $\rho_{xx} / \rho_{xx}(300\text{ K})$ as a function of temperature for the studied samples. Dashed lines are fits according to $\rho_{xx} = A + BT^n$. (b) $\rho_{xx}$ as a function of temperature at various temperatures for sample #4. (c) Magnetoresistance (MR) for several samples #2, #3, #5, and #6 at 2 K. Dashed lines are fits according to Eq. (4). Inset shows the MR and fits at 2 K for sample #1. (d) Quadrature scaling plot of $[\rho_{xx}(B,T) - \rho_{xx}(0,0)]/T$ vs $B/T$ for sample #2. Inset is an optical microscope image of sample #2.

Figure 2

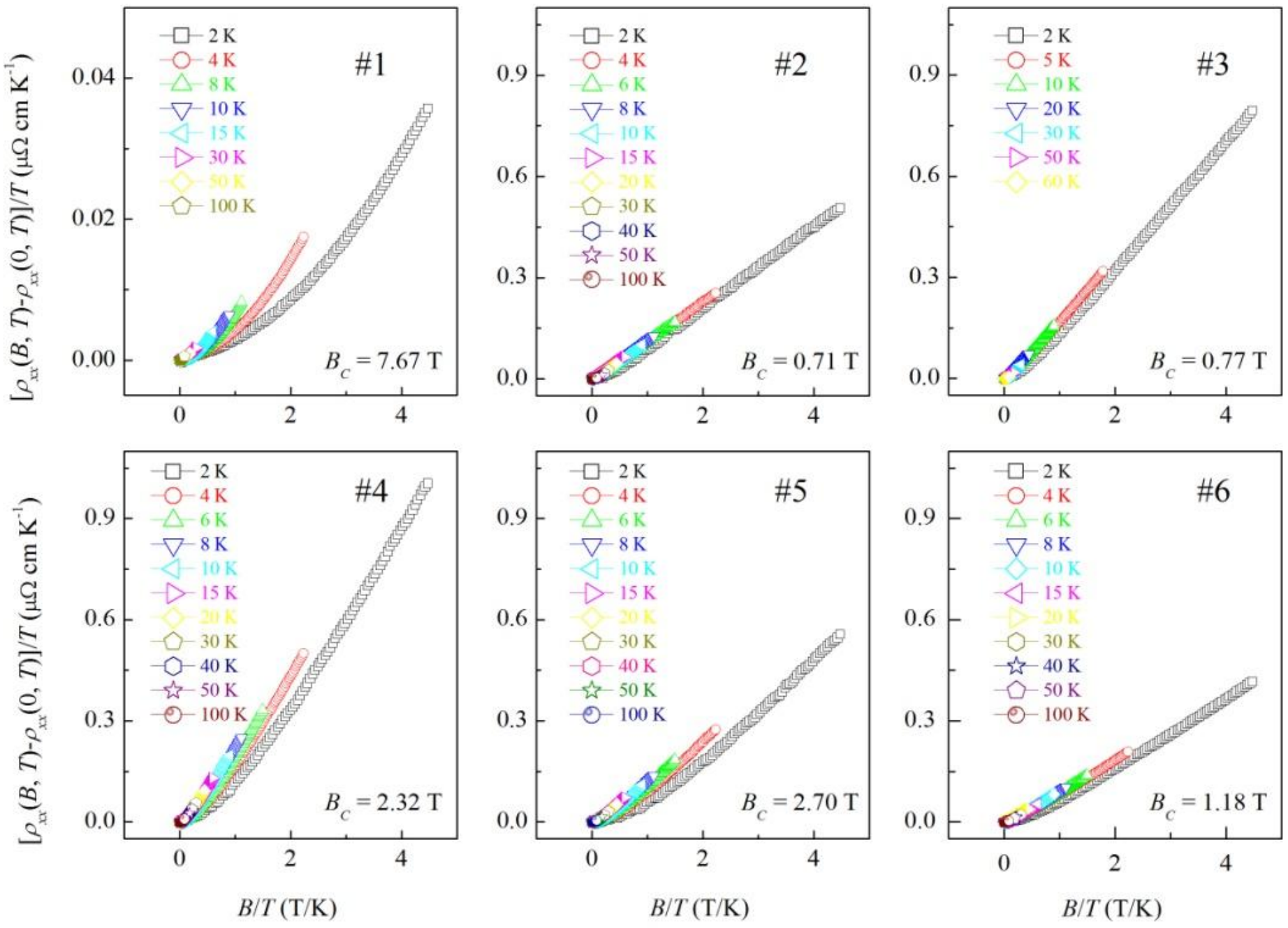

**Figure 2** Quadrature scaling plots of $[\rho_{xx}(B,T)-\rho_{xx}(0,T)]/T$ vs $B/T$ for six studied samples. The same range of *y*-coordinates is adopted for samples #2-#6. For clarity, a narrower range of *y*-coordinate is given for sample #1. $B_c$ value at 2 K is given for each sample.

Figure 3

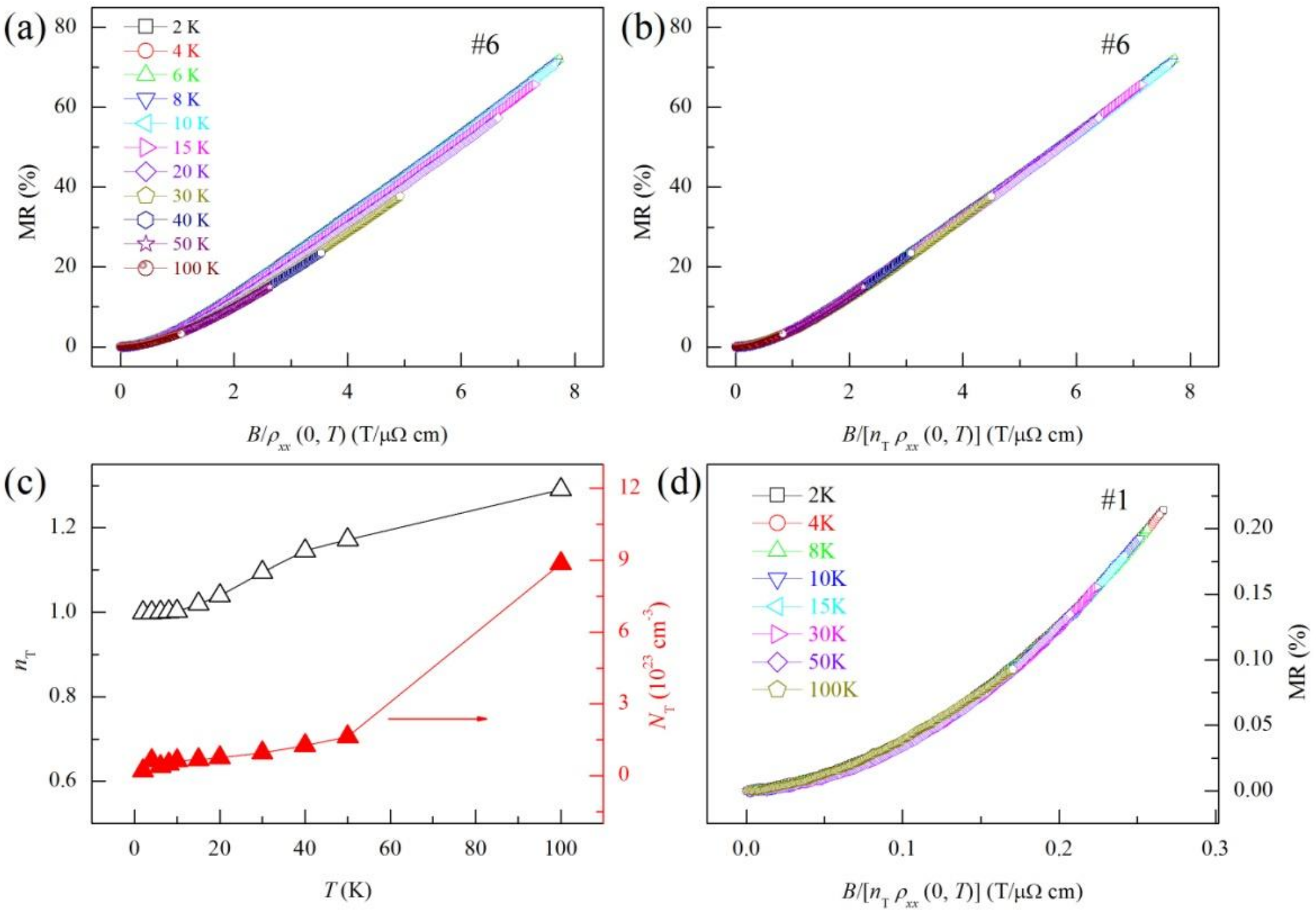

**Figure 3** (a) Kohler's rule plots of the MR for sample #6. (b) Extended Kohler's rule plots of the MR for sample #6. (c) The extracted $n_T$ and total carrier concentration $N_T$ as a function of temperature for sample #6. (d) Extended Kohler's rule plots of the MR for sample #1.

Figure 4

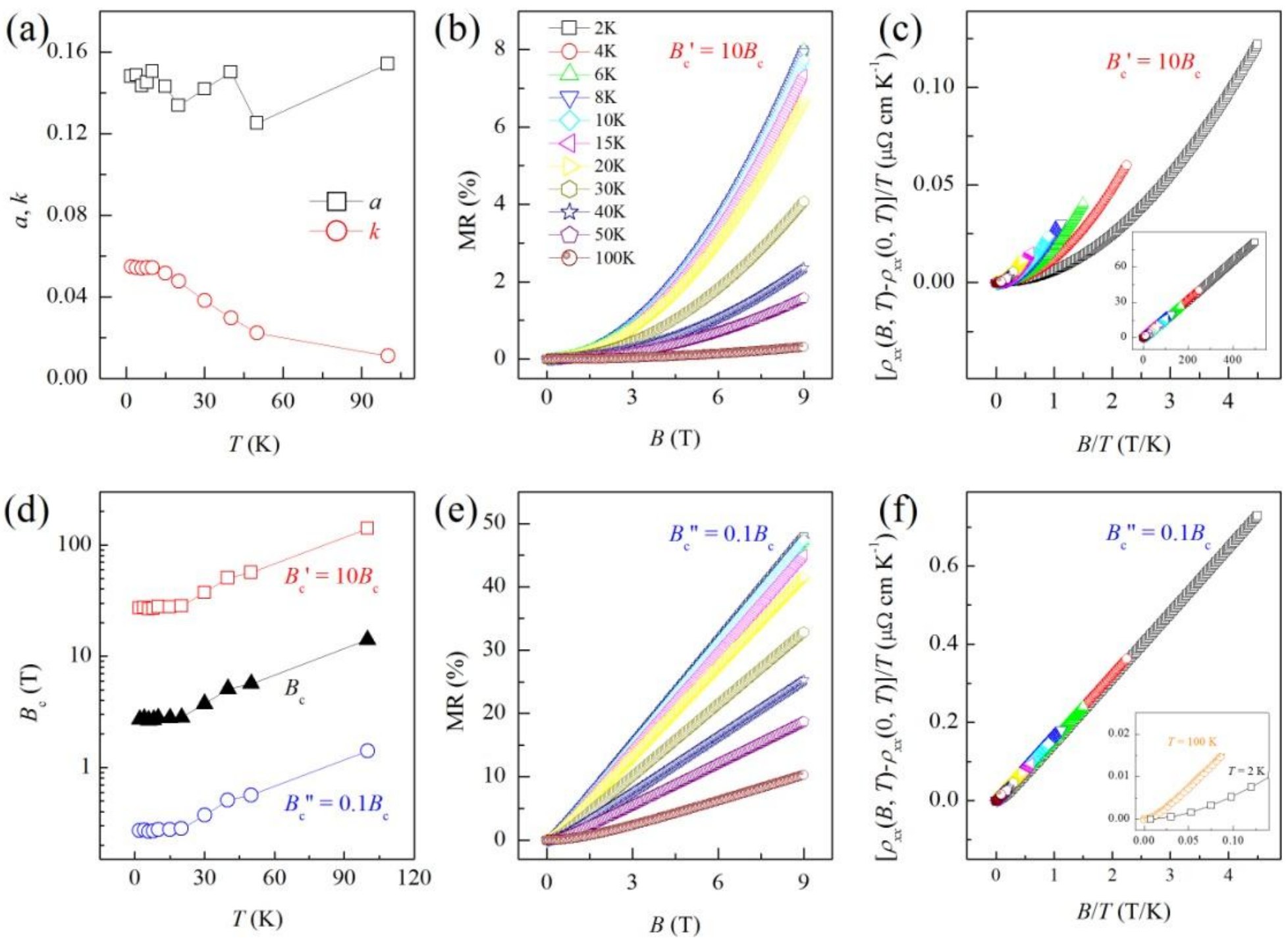

**Figure 4** (a) The extracted parameters $a$ and $k$ as a function of temperature for sample #5. (b) Simulated MR at various temperatures when $B_c' = 10B_c$. (c) Quadrature scaling plots for the same data in (b). Inset shows quadrature scaling plots in the wide field range of 0-1000 T when $B_c' = 10B_c$. (d) Crossover fields $B_c$ as a function of temperature for sample #5. $B_c'$, and $B_c''$ are two calculated crossover fields. (e) Simulated MR at various temperatures when $B_c'' = 0.1B_c$. (f) Quadrature scaling plots for the same data in (e). Inset is a comparison of two quadrature scaling plots with $T = 2$ K and 100 K in a low $B/T$ range.

Figure 5

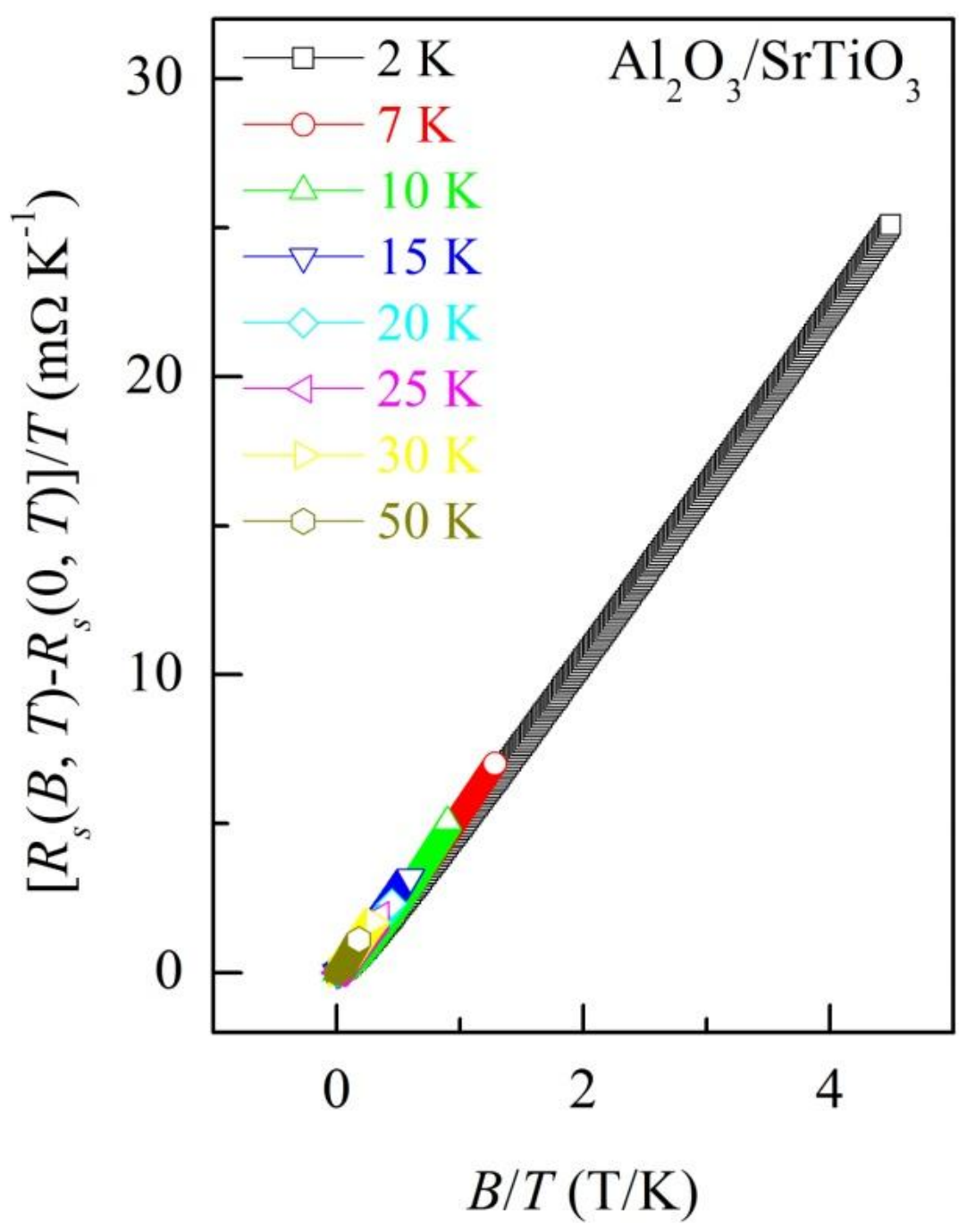

**Figure 5** Quadrature scaling plots of $[R_s(B,T)-R_s(0,T)]/T$ vs $B/T$ for $Al_2O_3/SrTiO_3$ heterostructures.

# Supplementary materials for:

# Quadrature magnetoresistance scaling reflects linear field dependence rather than strange metallicity of high-temperature superconductors

Dingbang Zhou,[1,2,*] Yang Yang,[1,*] Liefeng Feng,[1] Mengfan Zhao,[3] Zhiyan Jia,[3] and Kuanghong Gao[1,†]

[1]Tianjin Key Laboratory of Low Dimensional Materials Physics and Preparing Technology, Department of Physics, Tianjin University, Tianjin 300354, China

[2]Analysis and Testing Center of Tianjin University, Tianjin 300072, China

[3]Institute of Quantum Materials and Devices, Tiangong University, Tianjin 300387, China

[*]These authors have equal contributions to this work.

[†]Author to whom any correspondence should be addressed.

**E-mail:** khgao@tju.edu.cn

## 1. The power-law fit of $\rho_{xx}$ vs $T$ plots at low temperatures

As seen in Fig. 1(a), $\rho_{xx}$ exhibits a roughly quadratic temperature dependence for all samples at low temperatures below ~50 K (near saturation region). The relation $\rho_{xx} = A + BT^n$ was used to fit this dependence. During the fitting procedure, we found that the fit residual shows a large change for the different fitting ranges. As seen in Fig. S1 for a representative sample #4, the magnitude of the fit residual is gradually reduced when narrowing the fitting window. In the fitting range of 2-30 K, one can see that the fit residual is small enough (~0.01 μΩ cm), and the narrower fitting range (e.g., 2-20 K) cannot cover entire quadratic temperature dependence. We therefore adopt the fitting window of 2-30 K. Note that there is a small change in the temperature upper limit for different samples. In order to exhibit the evolution of the fitting curves at high temperatures, dashed lines in Fig. 1(a) is extrapolated up to 80 K. The extracted exponent $n$, as seen in Table 2, varies between 2.26±0.02 and 2.62±0.02.

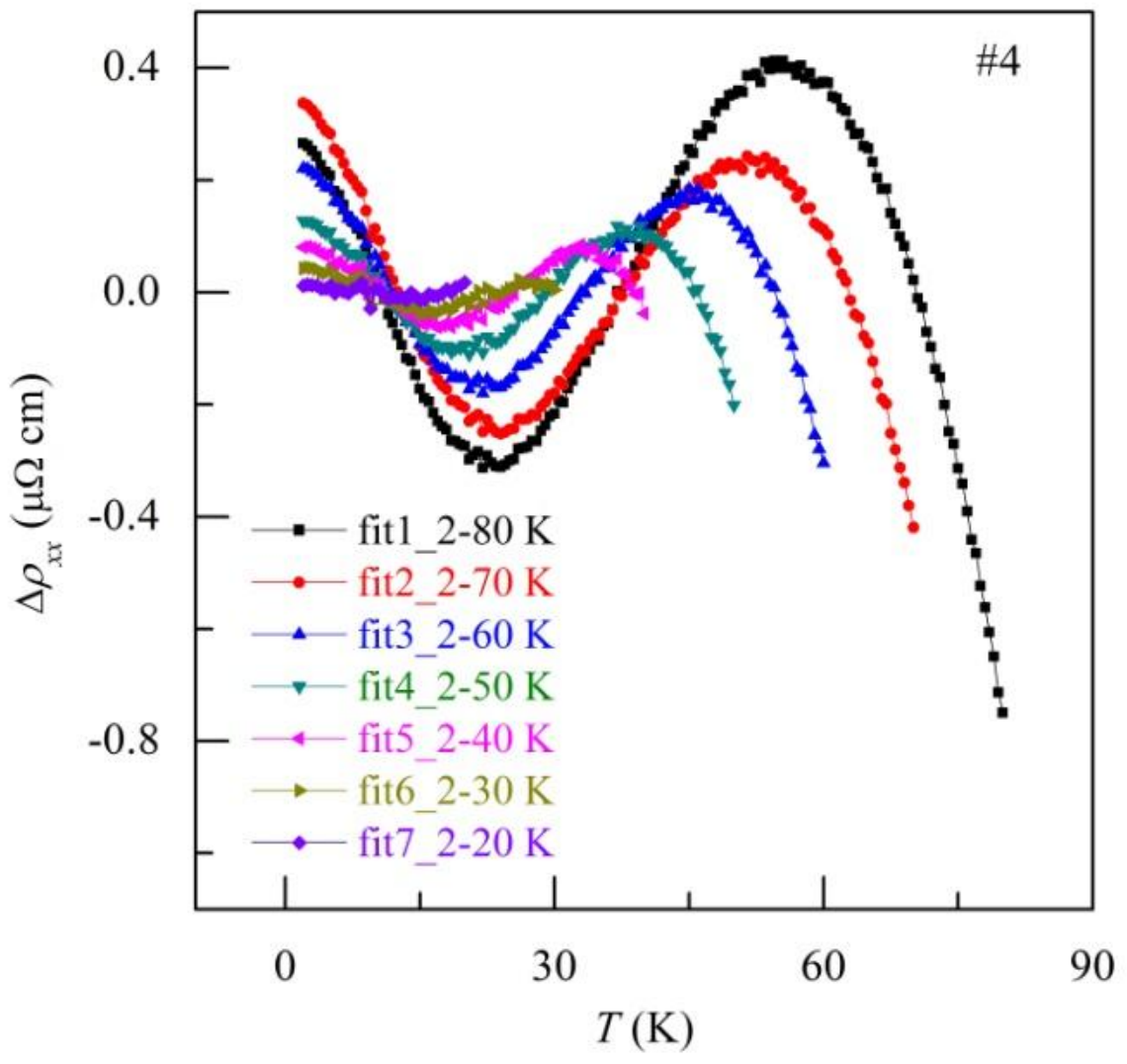

Supplementary Fig. S1 The fit residual $\Delta\rho_{xx} = \rho_{xx} - \rho_{\mathrm{fit}}$ by using relation $\rho_{xx} = A + BT^n$ to fit $\rho_{xx}$ vs $T$ curve at various fitting ranges for sample #4.

## 2. The Bloch-Grüneisen fit of $\rho_{xx}$ vs $T$ plots at high temperatures

$\rho_{xx}$ shows a quasi-linear temperature dependence at high temperatures where the electron-phonon scattering likely dominates. We try to apply the Bloch-Grüneisen formula [i.e., Eq. (2)] to fit the $\rho_{xx}$ vs $T$ plots in high temperature range. The different fitting windows are adopted [see Fig. S2(a) for sample #6]. The extracted parameters $\rho_{xx}(0,0)$, $M$, and $\theta_{\mathrm{D}}$ are shown in Fig. S2(b). It is found that both $\rho_{xx}(0,0)$ and $\theta_{\mathrm{D}}$ show an increase with narrowing the window while $M$ has no clear fitting-range dependence. Importantly, one can see that the extracted $\rho_{xx}(0,0)$ gradually deviates from the measured value (dashed arrow) with narrowing the fitting window. As a result, we adopted a wide the fitting window of 30-300 K. As shown in Fig. S2(c), the temperature dependence of $\rho_{xx}$ at high temperatures can be well described by Eq. (2) for all studied samples. Interestingly, as shown in Fig. S2(d), $\theta_{\mathrm{D}}$ roughly increases on increasing the thickness of the nanosheet.

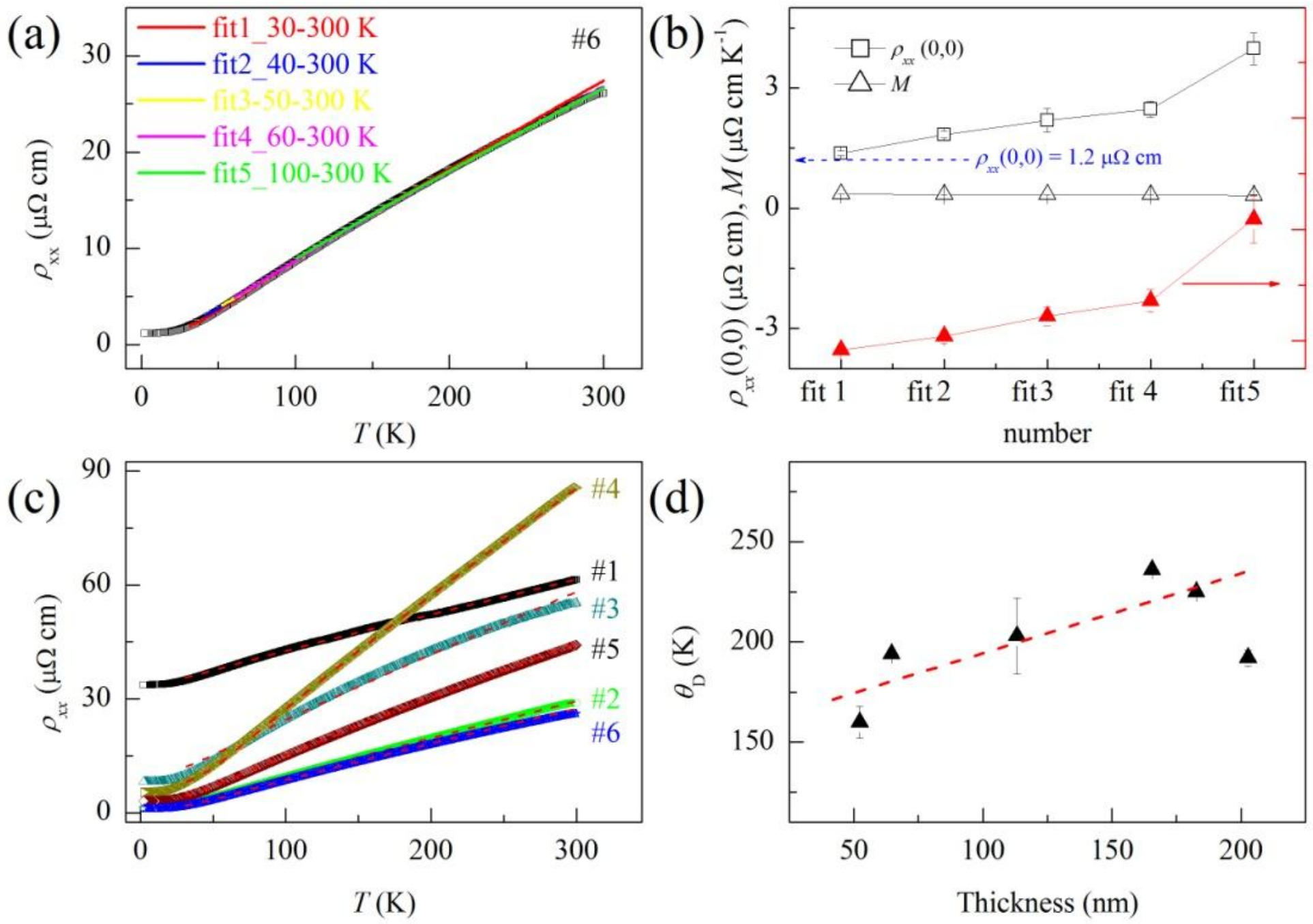

Supplementary Fig. S2 (a) $\rho_{xx}$ vs $T$ plot and the fits by using Eq. (2) at various fitting ranges for sample #6. (b) The fit-obtained parameters $\rho_{xx}(0,0)$, $M$ and $\theta_{\mathrm{D}}$ at various fitting ranges. Dashed arrow marks the experimentally measured $\rho_{xx}(0,0)$ value. (c) $\rho_{xx}$ vs $T$ plots and the fits by using Eq. (2) for all studied samples. Dashed lines are fits according to Eq. (2). (d) $\theta_{\mathrm{D}}$ as a function of thickness of nanosheet. Dashed line provides a guide for the eye.

## 3. Disorder-induced linear magnetoresistance (MR)

We find that the observed linear MR can be attributed to the classical disorder [19]. In the classical disorder model [39], unsaturated MR shows a quadratic field dependence at low fields and a linear field dependence at high fields. The magnitude (equivalently, the slope dMR/d$B$) of linear MR can be dominated either by carrier mobility or by its fluctuation. Namely, dMR/d$B$ is proportional to carrier mobility or its fluctuation. And the crossover field $B_{\mathrm{C}}$, marking a transition from quadratic to linear field dependence, is proportional to the inverse mobility. Apparently, one must obtain the carrier mobility for further analysis. To obtain the carrier mobility, the Hall data is studied.

As shown in Fig. S3(a) for a representative sample #2, Hall resistivity $\rho_{xy}$ vs $B$ plots are nonlinear for any given temperature, indicating a multiband occupation. A two-band model is adopted to analyze the data [S1]. In this model, $\rho_{xx}$ and $\rho_{xy}$ are given as

$$\rho_{xx}(B) = \frac{(n_{\mathrm{e}}\mu_e + n_{\mathrm{p}}\mu_p) + (n_{\mathrm{e}}\mu_p + n_{\mathrm{p}}\mu_e)\mu_e\mu_p B^2}{e[(n_{\mathrm{e}}\mu_e + n_{\mathrm{p}}\mu_p)^2 + (n_{\mathrm{p}} - n_{\mathrm{e}})^2 \mu_e^2\mu_p^2 B^2]}, \tag{S1}$$

$$\rho_{xy}(B) = \frac{(n_{\mathrm{p}}\mu_{\mathrm{p}}^2 - n_{\mathrm{e}}\mu_{\mathrm{e}}^2)B + (n_{\mathrm{p}} - n_{\mathrm{e}})\mu_{\mathrm{p}}^2\mu_{\mathrm{e}}^2 B^3}{e[(n_{\mathrm{e}}\mu_{\mathrm{e}} + n_{\mathrm{p}}\mu_{\mathrm{p}})^2 + (n_{\mathrm{p}} - n_{\mathrm{e}})^2 \mu_{\mathrm{e}}^2\mu_{\mathrm{p}}^2 B^2]}, \tag{S2}$$

where $n_{\mathrm{e}}$, $n_{\mathrm{p}}$, $\mu_{\mathrm{e}}$ and $\mu_{\mathrm{p}}$ are electron and hole concentrations and mobilities, respectively. We firstly tried to fit both $\rho_{xx}(B)$ and $\rho_{xy}(B)$ data simultaneously. Unfortunately, no satisfactory reproduction is obtained. It is understandable because Eq. (S1) only yields parabolic MR that differs from the observed linear MR at high fields [Fig. 1(c)]. We thus only fit $\rho_{xy}(B)$ by using Eq. (S2), and the same procedure was widely reported [S2,S3]. One can see that $\rho_{xy}(B)$ data can be well described by Eq. (S2) in Fig. S3(a) (dashed lines are fitting curves). The extracted $\mu_{\mathrm{e}}$ and $\mu_{\mathrm{p}}$ vary between 29 and 2670 $\mathrm{cm^2V^{-1}s^{-1}}$, and the extracted $n_{\mathrm{e}}$ and $n_{\mathrm{p}}$ are in the order of $10^{21}$-$10^{24}$ $\mathrm{cm^{-3}}$, which are comparable to the reported values in $NiTe_2$ [18,36]. According to the extracted parameters, one can obtain the total carrier concentration $N_{\mathrm{T}}$ by using $N_{\mathrm{T}} = n_{\mathrm{e}} + n_{\mathrm{p}}$ and the effective mobility $\mu_{\mathrm{E}}$ by using $\mu_{\mathrm{E}} = (n_{\mathrm{e}}\mu_{\mathrm{e}} + n_{\mathrm{p}}\mu_{\mathrm{p}})/N_{\mathrm{T}}$. It should be mentioned that the ratio $n_{\mathrm{p}}/n_{\mathrm{e}}$ varies between 1.1 and 17.4 for all studied samples at 2 K. This means that the carrier transport deviates from the perfect carrier compensation (i.e., $n_{\mathrm{p}}/n_{\mathrm{e}} = 1$), which rules out the proposed mechanism of compensation-caused linear MR [45].

As seen in Fig. S3(b), dMR/d$B$ in the linear field region is proportional to $\mu_{\mathrm{E}}$ (i.e., $\mathrm{dMR}/\mathrm{d}B \propto \mu_{\mathrm{E}}$) for sample #2, consistent with the prediction of the classical disorder model. Fig. S3(c) shows the fit-obtained $B_{\mathrm{C}}$ by using Eq. (4) at various temperatures as a function of $\mu_{\mathrm{E}}^{-1}$. One can see that $B_{\mathrm{C}}$ is proportional to $\mu_{\mathrm{E}}^{-1}$ (i.e., $B_C \propto \mu_E^{-1}$), confirming that the observed linear MR originates from the disorder. On the other hand, as seen in Fig. S4(a), the dMR/d$B$ of MR curves at 2 K for different samples is also proportional to $\mu_E$, in agreement with the classical disorder model. However, as seen in Fig. S4(b), $B_C$ deviates from the relation $B_C \propto \mu_E^{-1}$ for samples with higher $\mu_E$ (i.e., #4-#6). This can be explained by the presence of multiband effect-induced parabolic MR that becomes important in samples with higher mobility [S1]. The multiband effect-induced parabolic MR may be comparable to the disorder-induced quadratic MR in these samples, and thus makes $B_C$ larger (consequently cause a deviation). Indeed, as seen in Fig. S4(c) for a representative sample #5 with the higher mobility, it is found that $B_C$ at low temperatures (blue dashed line marks) also deviates from the relation $B_C \propto \mu_E^{-1}$ because $\mu_E$ is higher in this temperature region. It can be concluded that $B_C$ is controlled dominantly by the classical disorder while the multiband effect becomes important to determine $B_C$ value in the low temperature region for high-mobility samples like #5.

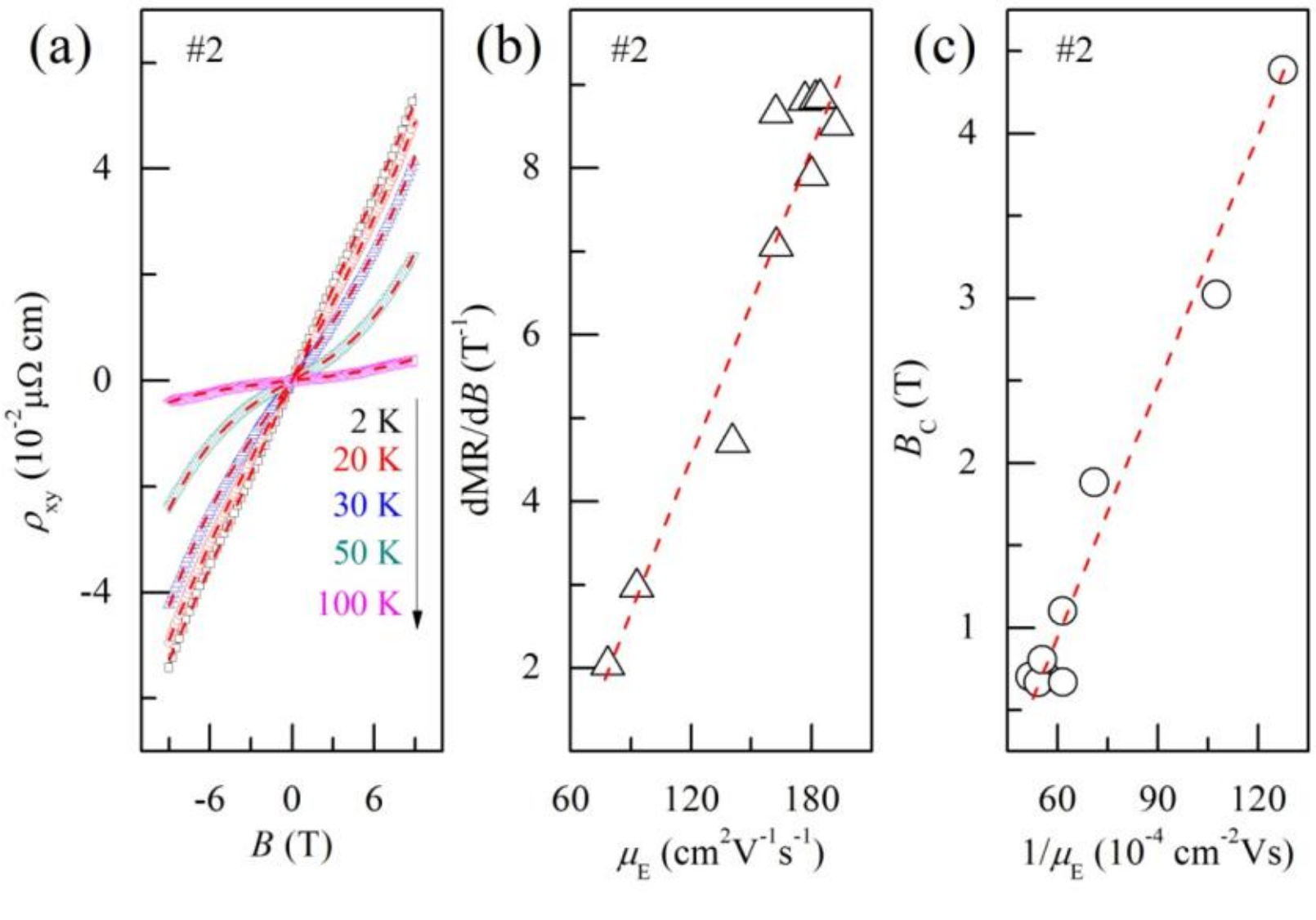

Supplementary Fig. S3 (a) Hall resistivity $\rho_{xy}$ as a function of field at various temperatures for sample #2. Dashed lines are fits according to Eq. (S1). (b) First derivative of linear MR, i.e.,dMR/d$B$, in high-field range as a function of $\mu_E$ at 2 K. Dashed line is a guide to the eye. (c) The crossover field $B_C$ as a function of the inverse of $\mu_E$. Dashed line is a guide to the eye.

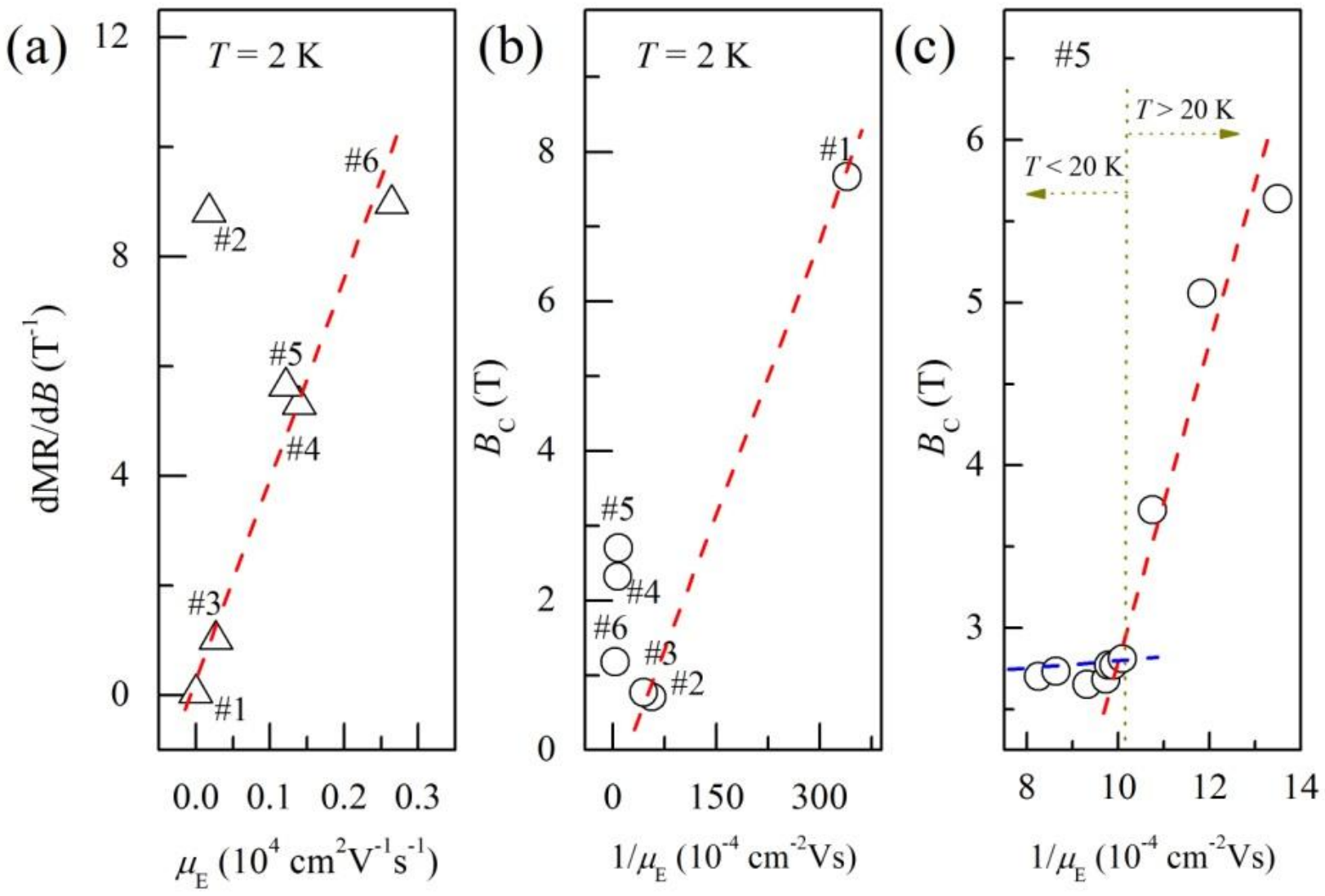

Supplementary Fig. S4 (a) dMR/d*B* as a function of $\mu_E$ at 2 K for all samples. Dashed line is a guide to the eye. (b) $B_C$ as a function of the inverse of $\mu_E$ at 2 K for all samples. Dashed line is a guide to the eye. (c) $B_C$ as a function of the inverse of $\mu_E$ for sample #5. Red (blue) dashed line provides the guide to the eye in the high (low) temperature range. Dotted line marks the data point at 20 K.

## 4. Origin of the approximate quadrature scaling above ~8 K for samples #4 and #5

Since the quadrature scaling only captures the linear MR character, a perfect collapse of $[\rho_{xx}(B,T)-\rho_{xx}(0,T)]/T$ vs $B/T$ plots at different temperatures for a given sample needs two conditions: (1) the same slope of linear MR (i.e., dMR/d*B*) and (2) the same $B_c/T$ value for different temperatures.

As seen in Fig. S5(a), dMR/d*B* is temperature independent in low temperatures and starts to decrease above ~20 K for both samples #4 and #5. Considering the aforementioned relation $\mathrm{dMR}/\mathrm{d}B \propto \mu_E$, the decrease of the dMR/d*B* can be attributed to the lower mobility at high temperatures plausibly due to electron-phonon scattering that also cause an increase of $\rho_{xx}$ [see Fig. (a)]. On the other hand, as seen in Fig. S5(b), the $B_c/T$ for both samples shows a rapid decrease with increasing temperature from 2 K up to ~20 K. But above ~20 K, one can find that the $B_c/T$ is nearly temperature independent for both samples. It is concluded that there is no significant change near 8 K for both dMR/d$B$ and $B_c/T$ values. And the approximate quadrature scaling above 8 K for both samples #4 and #5 in Fig. 2 is not related directly to the quasi-linear temperature dependent $\rho_{xx}$ above ~30 K [see Fig. 1(a)].

Since there is no a fixed $B_c/T$ value at low temperatures for theses two samples, it is understandable that the quadrature scaling fails below 8 K. The approximate quadrature scaling above 8 K for both samples #4 and #5 in Fig. 2 can be attributed to the narrow $B/T$ window of 0-1.1 T/K compared to the maximum of 4.5 T/K, which can be named after the window effect. Indeed, as shown in Fig. S5(c) for $[\rho_{xx}(B,T)-\rho_{xx}(0,T)]/T$ vs $B/T$ plots in the temperature range of 8-100 K, the quadrature scaling fails. Notably, the approximate quadrature scaling appears again in the higher temperature range of 20-100 K in Fig. S5(c). This must arise from the temperature independence of the $B_c/T$ [see orange solid line Fig. S5(b)] while the varying dMR/d$B$–caused the deviation of the collapse is overwhelmed by the window effect.

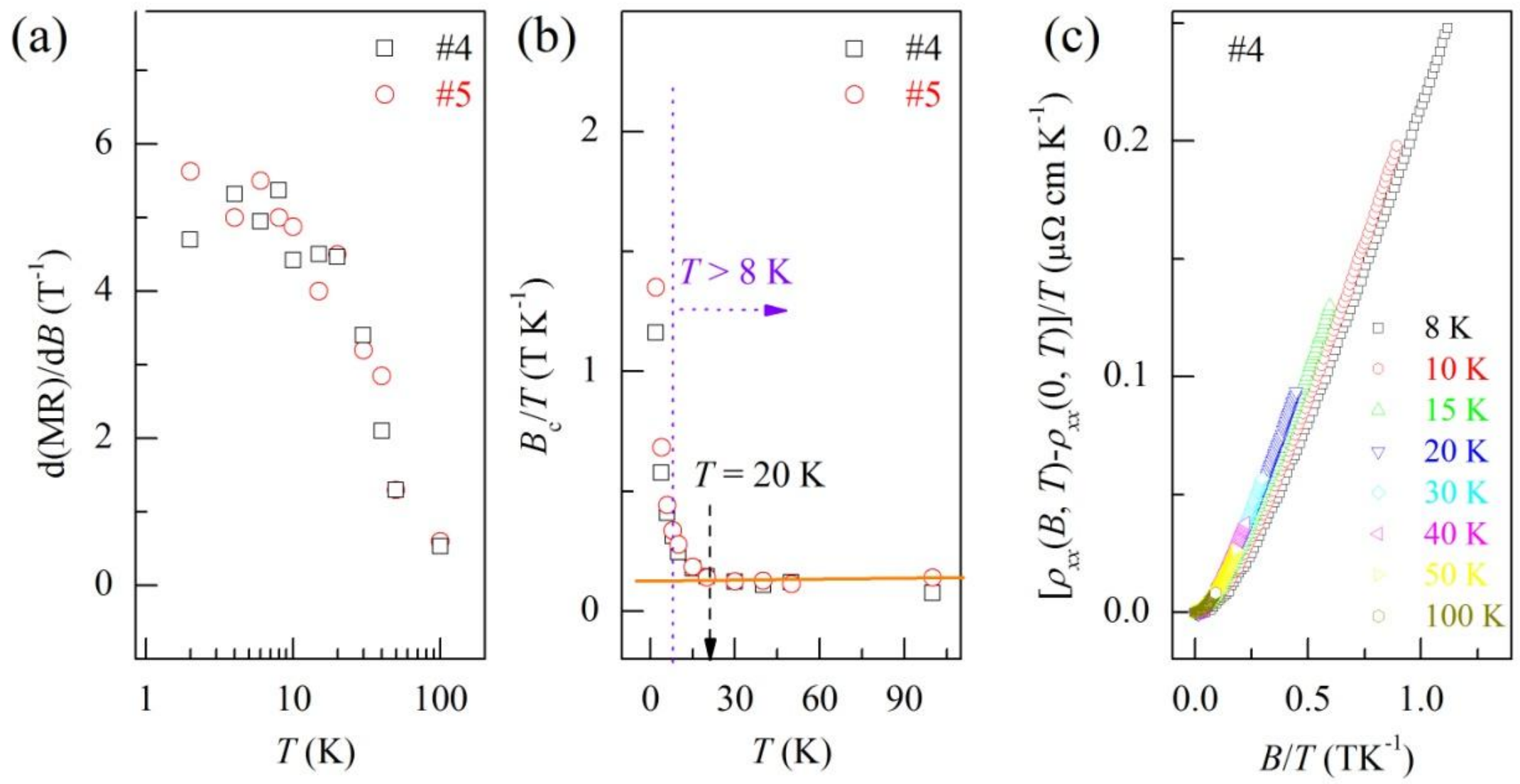

Supplementary Fig. S5 (a) dMR/d$B$ as a function of $T$ for two samples #4 and #5. (b) $B_c/T$ as a function of $T$ for samples #4 and #5. Orange solid line is a guide to the eye. (c). Quadrature scaling plots of $[\rho_{xx}(B,T)-\rho_{xx}(0,T)]/T$ vs $B/T$ in the temperature range of (8-100 K) for #4.

## 5. $[\rho_{xx}(B,T)-\rho_{xx}(0,0)]/T$ vs $B/T$ plots of the quadrature scaling

Fig. S6 shows the $[\rho_{xx}(B,T)-\rho_{xx}(0,0)]/T$ vs $B/T$ plots for all studied samples in the low temperature range of 2-20 K. One can find the same phenomenon: the quadrature scaling hold in samples #2, #3, and #6 but fail in others, in agreement with $[\rho_{xx}(B,T)-\rho_{xx}(0,T)]/T$ vs $B/T$ plots shown in Fig. 2. This indicates that there is no remarkable difference when $\rho_{xx}(0,0)$ is substituted by $\rho_{xx}(0,T)$ for the quadrature scaling at low temperatures. In Fig. 2, one can note that those $[\rho_{xx}(B,T)-\rho_{xx}(0,T)]/T$ vs $B/T$ plots above 20 K are within a narrow $B/T$ range of 0-0.5 T/K. As a consequence, those plots above 20 K have a negligible influence on the quadrature scaling.

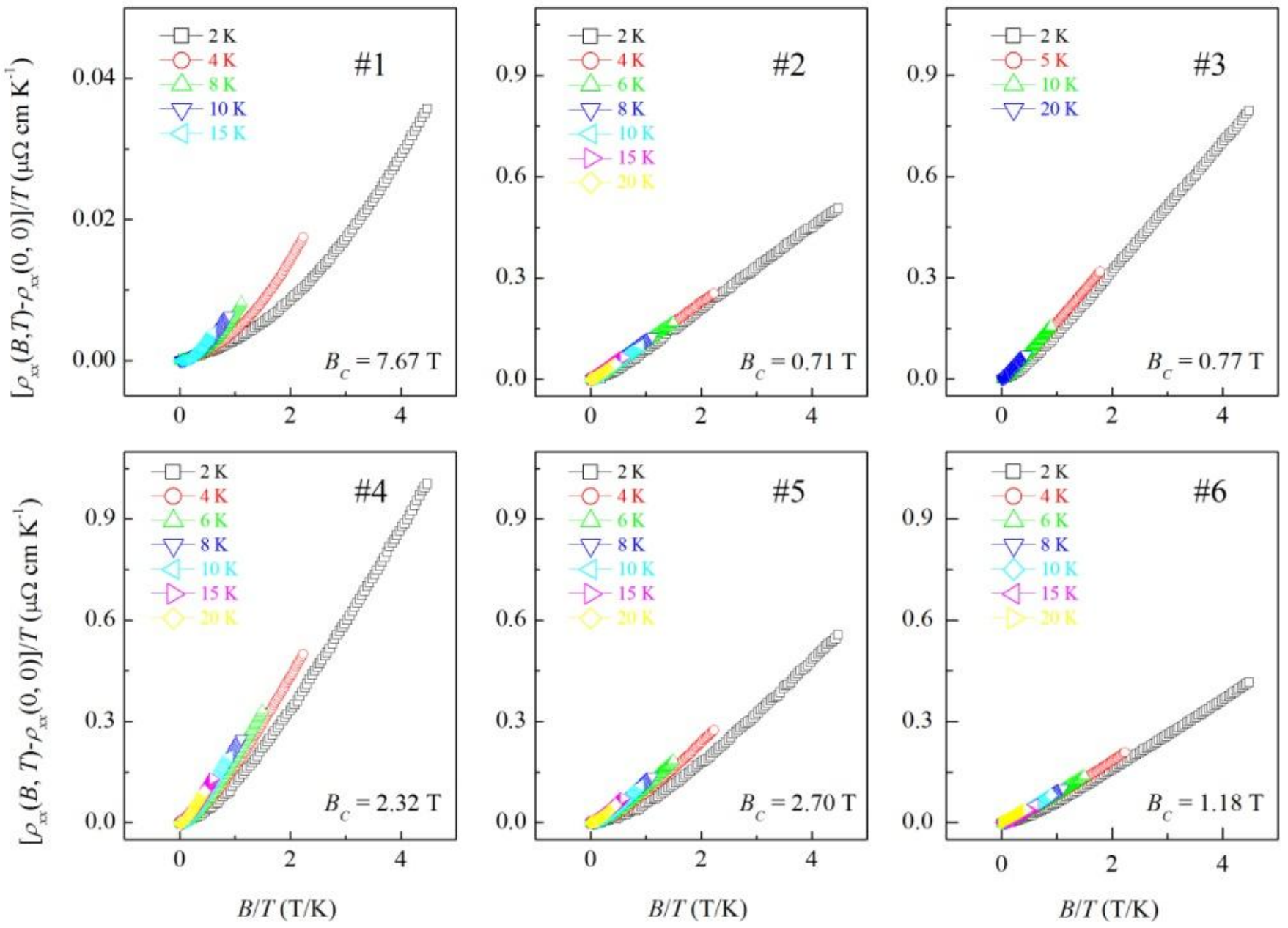

Supplementary Fig. S6 Quadrature scaling plots of $[\rho_{xx}(B,T)-\rho_{xx}(0,0)]/T$ vs $B/T$ for all studied samples at low temperatures (2-20 K).

**Supplementary References**